\documentclass[conference]{IEEEtran}
\IEEEoverridecommandlockouts
	\usepackage[turkish]{babel}
	\usepackage[utf8]{inputenc} % Kullanılan encodinge göre utf8 yerine latin5 de yazılabilir.
	\usepackage[T1]{fontenc}

\usepackage{cite}
\ifCLASSINFOpdf
  \usepackage[pdftex]{graphicx}
\else
\fi
\usepackage[cmex10]{amsmath}
\usepackage{multirow}
\usepackage{array}
\usepackage[lofdepth,lotdepth]{subfig}
\usepackage{lipsum}
\usepackage{mathtools}
\usepackage{cuted}
\usepackage{xfrac}

\AtBeginDocument{%
  \renewcommand\tablename{TABLO}
}

\AtBeginDocument{%
  \renewcommand\abstractname{Abstract}
}

\begin{document}

\IEEEpubid{\makebox[\columnwidth]{978-1-5386-1501-0/18/\$31.00 \copyright\ 2018 IEEE\hfill}
\hspace{\columnsep}\makebox[\columnwidth]{}}

\title{Nakagami-m Sönümlemeli Kanallarda DL-NOMA için Kapalı-Form BHO İfadelerinin Türetilmesi \\
Derivation of the Closed-Form BER Expressions for DL-NOMA over Nakagami-m Fading Channels}

\author{\IEEEauthorblockN{\textit{Ferdi KARA, Hakan KAYA}}
\IEEEauthorblockA{Elektrik-Elektronik Mühendisliği
\\ Bülent Ecevit Üniversitesi
\\Zonguldak, Türkiye
\\ \{f.kara, hakan.kaya\}@beun.edu.tr
}
}
\maketitle
\begin{ozet}
NOMA kitlesel iletişimi destekleyebilme potansiyeli ve yüksek spektral verimlilik sunması sayesinde gelecek nesil Radyo Erişim Ağları (FRA) için önemli bir aday tekniktir. Fakat, kullanıcılar arası girişim sebebiyle alıcılarda gerçekleştirilen Ardışık Girişim Engelleyici (SIC) sırasında oluşan hatalar NOMA’nın en büyük dezavantajı olarak gözükmektedir. Bu çalışmada, literatürde ilk defa SIC hatalarının varlığında Nakagami-$\pmb{m}$ sönümlemeli kanallar üzerinde aşağı yönlü (DL) NOMA için Bit-Hata Oranı ifadeleri kapalı formda türetilmiştir.  Türetilen ifadeler bilgisayar simülasyonları ile desteklenmiştir. NOMA’da da tıpkı OMA sistemlerinde olduğu gibi, $\pmb m$ parametresinin çeşitlilik derecesini ifade ettiği gözlenmiştir. Buna ek olarak, güç paylaşım katsayısı da NOMA kullanıcılarının BHO başarımları üzerinde önemli bir etkiye sahiptir. 
%\boldmath
\end{ozet}
\begin{IEEEanahtar}
NOMA, BHO, SIC hataları.
\end{IEEEanahtar}
\begin{abstract}
NOMA is as a strong candidate for the Future Radio Access Network (FRA) due to its potential to support massive connectivity and high spectral efficiency. However, the most important drawback of NOMA is the error during Successive Interference Canceller (SIC) is implemented because of the inter-user interferences. In this paper, we derive closed-form exact Bit-Error Rate expressions for Downlink(DL) NOMA over Nakagami-$\pmb{m}$ fading channels in the presence of SIC errors. The derived expressions are validated by the computer simulations. It is shown that the $\pmb{m}$ parameter still represents the diversity order like as OMA systems. Besides, the BER performances of users for NOMA have substantially depended on the power allocation coefficient. 
%\boldmath
\end{abstract}
\IEEEpeerreviewmaketitle
\IEEEpubidadjcol
\section{G{\footnotesize İ}r{\footnotesize İ}ş}
Dikgen Olmayan Çoklu Erişim (Non-Orthogonal Multiple Access -NOMA) kitlesel iletişime olanak tanıması ve yüksek spektral verimliliği nedeniyle 5G ve sonrası kablosuz sistemler için önemli bir aday olarak öne çıkmaktadır [1]. NOMA bu potansiyeli sebebiyle son yıllarda akademiden ve endüstriden birçok araştırmacının ilgisini çekmektedir [2]-[5]. NOMA temelde iki veya daha fazla kullanıcının aynı radyo kaynaklarını (zaman, frekans) kullanması prensibine dayanmaktadır. Geleneksel Dikgen Çoklu Erişim (Orthogonal Multiple Access- OMA) sistemlerinde farklı kullanıcılara ait semboller dikgen zaman ya da frekans kaynaklarını kullanarak alıcılara iletilir. NOMA’da ise bu işlem kullanıcıların yeni bir eksen (domain) olarak güç (power) ekseninde bölünmesiyle sağlanır. Aşağı yönlü (Downlink -DL) iletişimde bu bölme süperpozisyon (superposition -SC) kodlaması yapılarak sağlanır. Baz istasyonunda SC kullanılarak tüm kullanıcıların sembolleri farklı güç bileşenleri ile çarpılarak toplanır ve alıcılara aynı radyo kaynaklarını kullanarak gönderilir. SC işlemi sırasında toplam güç verici ile alıcılar arasındaki kanal katsayılarının azalan sırasına göre paylaştırılır. Böylece en kötü kanala sahip kullanıcıya en yüksek güç aktarılmış olur. Bu durum geleneksel OMA sistemlerine göre kullanıcılar arası daha adil (user fairness) bir iletişim imkanı sunmaktadır [6]. Alıcılarda (gezgin kullanıcılar) ise kullanıcılar arası girişimi yok etmek için Ardışık Girişim Engelleyici (Successive Intereference Canceller - SIC) kullanılır. SIC işleminde alıcılardan her biri öncelikli olarak çözme sırasına (decoding order) göre yüksek güç bileşenine sahip kullanıcıların verilerini sezer ve bu sezilen işaretleri toplam alınan işaretten çıkararak kendi sembollerini elde ederler. NOMA’nın kesinti olasılığı, toplam kapasite/hız açısından OMA tekniklerine üstünlük sağladığı gösterilmiştir [7],[8]. Fakat literatürde yapılan çalışmaların çoğunda, SIC sırasında yapılan hatalar göz ardı edilmiştir. Bu varsayım kablosuz haberleşme teknikleri için makul bir yaklaşım değildir. Diğer taraftan, NOMA’nın Bit Hata Oranı (BHO, Bit Error Rate BER) ifadelerinin çıkarımı ile ilgili yapılan çalışmaların çoğunluğu sadece benzetim sonuçlarını içermektedir ve analitik çözümler bulunmamaktadır [9],[10]. Analitik çözümlerin bulunduğu çalışma olan [11]'de ise, yukarı yönlü (Uplink) iletişim için kullanıcılar ile baz istasyonu arasındaki kanal katsayıların sabit olduğu varsayılıp çözümlemeler Toplanabilir Beyaz Gauss Gürültülü (Additive White Gaussian Noise- AWGN) kanallar için sunulmuştur. Kanal sönümlemeleri göz ardı edilmiştir. 

Bu çalışmada DL-NOMA için BHO ifadeleri sönümlemeli kanallarda analitik olarak kapalı formda sunulmuştur. Elde edilen analitik sonuçlar bilgisayar benzetimleri ile desteklenmiştir. Bu çalışmanın II. Bölümünde sistem modeli tanıtılmıştır. III. Bölümde, türetilen BHO analitik ifadeleri kapalı-formda verilmiştir.  Daha sonra, elde edilen analitik ifadelerin, bilgisayar benzetimleri ile doğrulanması IV. Bölümde sunulmuştur. Son olarak, V. Bölümde sonuçlar tartışılarak çalışma sonlandırılmıştır.
\section{S{\footnotesize İ}stem Model{\footnotesize İ}}
Bu çalışmada DL NOMA için iki adet gezgin kullanıcının ve bir baz istasyonunun bulunduğu sistem ele alınmıştır. SC işlemi sırasında toplam verici gücü $\ (P_s)$, $\alpha$ güç paylaşım katsayısı (power allocation coefficient) ile kullanıcıların sembolleri arasında paylaştırılmaktadır. Sistemde tüm düğümlerin tek antene sahip olduğu düşünülmektedir. Baz istasyonu ile kullanıcılar arasındaki kanal kaliteleri için $\left|h_1\right|^2>\left|h_2\right|^2$ varsayımı yapılırsa; 1. kullanıcının hücre içi (yakın) kullanıcı 2. kullanıcının ise hücre kenarı (uzak) kullanıcı olduğu düşünülebilir. Bu durumda çözme sırasında (decoding order) 2. kullanıcı önce gelmektedir. Bu nedenle 2. kullanıcıda yakın kullanıcı sembollerine gürültü gibi davranılarak klasik En Büyük Olabilirlikli (Maximum Likelihood ML) sezme işlemi gerçekleştirilecektir. Yakın kullanıcı olan 1. kullanıcıda ise SIC işlemi gerçekleştirilmelidir. Kullanıcılara ait modülasyon seviyeleri kullanıcıların baz istasyonu ile aralarındaki kanal kalitelerine göre belirlenmiştir. Uzak kullanıcı sembolleri için İkili Faz Kaydırmalı Anahtarlama (Binary Phase Shift Keying -BPSK) kullanılırken yakın kullanıcı sembolleri için görece daha yüksek seviyeli Dördün Faz Kaydırmalı Anahtarlama (Quadrature Phase Shift Keying -QPSK) kullanılmıştır. Gezgin kullanıcılar tarafından toplam alınan işaret 
\begin{equation}
\ y_k=h_k\sqrt{\varepsilon_1}x_1+h_k\sqrt{\varepsilon_2}x_2+w_k\\ \\ \ \ k=1,2
\end{equation}
şeklinde ifade edilebilir. Burada $\varepsilon_1=\alpha P_s$, $\varepsilon_2=\left(1-\alpha\right)P_s$ şeklinde ve $x_1$, $x_2$ sırasıyla QPSK ve BPSK modülasyonlu semboller olarak verilir. $h_k$ baz istasyonu ile $k$. kullanıcı arasındaki Nakagami-$m$ sönümlemeli kanal katsayısı ve $w_k$ da $k$. kullanıcı alıcısında oluşan sıfır “0” ortalamalı ve $\sfrac{N_0}{2}$ varyansa sahip AWGN olarak tanımlanır.
\section{BHO Anal{\footnotesize İ}zler{\footnotesize İ}}
Baz istasyonunda gerçekleştirilen SC işlemi sonrasında vericide oluşan toplam karmaşık işaret uzayı gönderilen farklı $x_2$ ve $x_2$ sembolleri için [11]'e benzer şekilde Tablo I'de verilmiştir. İkili bit (binary) gösterimlerde birinci alt indis kullanıcıyı ikinci alt indis ise bit sırasını ifade etmektedir.
\begin{table}[ph]
  \centering
  \caption{\textsc{Gönder{\footnotesize İ}len Sembollere Göre Ver{\footnotesize İ}c{\footnotesize İ}dek{\footnotesize İ} Karmaşık {\footnotesize İ}şaret Uzayı}}
  \label{tablo1}
  \begin{tabular}{|c|c|c|}
    \hline 
$\{x_1\}\backslash \{x_2\}$&$b_2=0$&$b_2=1$\\
	\hline
	$b_{1,2}b_{1,1}=00$&$\{-\sqrt{\frac{\varepsilon_1}{2}}-\sqrt{\varepsilon_2},-i\sqrt{\frac{\varepsilon_1}{2}}\}$&$\{-\sqrt{\frac{\varepsilon_1}{2}}+\sqrt{\varepsilon_2},-i\sqrt{\frac{\varepsilon_1}{2}}\}$\\
		\hline
	 $b_{1,2}b_{1,1}=01$&$\{-\sqrt{\frac{\varepsilon_1}{2}}-\sqrt{\varepsilon_2},i\sqrt{\frac{\varepsilon_1}{2}}\}$&$\{-\sqrt{\frac{\varepsilon_1}{2}}+\sqrt{\varepsilon_2},i\sqrt{\frac{\varepsilon_1}{2}}\}$\\	
		\hline
	$b_{1,2}b_{1,1}=10$&$\{\sqrt{\frac{\varepsilon_1}{2}}-\sqrt{\varepsilon_2},-i\sqrt{\frac{\varepsilon_1}{2}}\}$&$\{\sqrt{\frac{\varepsilon_1}{2}}+\sqrt{\varepsilon_2},-i\sqrt{\frac{\varepsilon_1}{2}}\}$\\
			\hline
		$b_{1,2}b_{1,1}=11$&$\{\sqrt{\frac{\varepsilon_1}{2}}-\sqrt{\varepsilon_2},i\sqrt{\frac{\varepsilon_1}{2}}\}$&$\{\sqrt{\frac{\varepsilon_1}{2}}+\sqrt{\varepsilon_2},i\sqrt{\frac{\varepsilon_1}{2}}\}$\\
		\hline
  \end{tabular}
\end{table}

Yakın kullanıcı ve uzak kullanıcı sembollerinin kendi aralarında eşit öncül olasılıklara sahip olduğu varsayılırsa, uzak kullanıcıda gerçekleştirilen ML sezici için karar kuralı $y_{2,R}<0$ veya $y_{2,R}\geq0$ olarak tanımlanabilir. Burada $y_{2,R}$ uzak kullanıcıda alınan işaretin gerçel bileşenini ifade etmektedir. Uzak kullanıcı sembolleri BPSK modülasyonlu işaretler olduğu için alınan işaretin sanal bileşeninin $(y_{2,I})$ karar kuralına herhangi bir etkisi yoktur.  Denklem (1)’ deki sönümleme katsayısı ve AWGN ile Tablo I’de verilen işaret uzayı göz önünde alındığında uzak kullanıcı için BHO ifadesi
\begin{equation}
\begin{split}
P_2\left(e\right)=&0.5\times\left[P_r\left(n_R\geq\sqrt{\varepsilon_2}h_2+\sqrt{\frac{\varepsilon_1}{2}}h_2\right) \right.\\
&\left.+P_r\left(n_R\geq\sqrt{\varepsilon_2}h_2-\sqrt{\frac{\varepsilon_1}{2}}h_2\right)\right]
\end{split}
\end{equation}
şeklinde ifade edilebilir. Burada $n_R$ AWGN’nin gerçel bileşenini göstermektedir. Bu durumda;
\begin{equation}
P_2\left(e\right)=0.5\times\left[Q\left(\sqrt{\gamma_A}\right)+Q\left(\sqrt{\gamma_B}\right)\right]
\end{equation}
olarak bulunur. Gösterim basitliği açısından $\gamma_A=\frac{\left(\sqrt{2\varepsilon_2}+\sqrt{\varepsilon_1}\right)^2\times\left|h_2\right|^2}{N_0}$ ve $\gamma_B=\frac{\left(\sqrt{2\varepsilon_2}-\sqrt{\varepsilon_1}\right)^2\times\left|h_2\right|^2}{N_0}$  olarak tanımlanır. [12, Denklem (9)]'da verilen alternatif $Q $ işlevi ve $h_2$ katsayılarının Nakgami-m dağılımlı olduğu durumda, [13, Denklem (17)] de verilen Moment Üreten Fonksiyon ifadeleri kullanılarak [14, Denklem (5.17)] verilen ifade yardımıyla  uzak kullanıcı için ortalama BHO ifadesi
 \begin{equation}
  \overline{P_2}\left(e\right) = \begin{cases}
        \frac{1}{2}\left[I\left(c_A\right)-I\left(c_B\right)\right] \quad\text{$m$ tamsayı,}
				\\
        \frac{1}{2}\left[J\left(c_A\right)-J\left(c_B\right)\right] \quad\text{diğer}.
        \end{cases}
 \end{equation}
olarak bulunur. Burada $c_i=\frac{\overline{\gamma_i}}{2m},\ i=A,B$, $\overline{\gamma_A}=E\left[\gamma_A\right]$ ve $\overline{\gamma_B}= E\left[\gamma_B\right]$ olarak tanımlanmaktadır. $E\left[.\right]$ beklenen değer işlevini gösterir. Denklem (4)’ de verilen $I\left(.\right)$ ve $J\left(.\right)$ işlevleri 
\begin{equation}
\begin{split}
\ &I\left(c\right)=\frac{1}{2}\left[1-\mu\left(c\right)\sum_{k=0}^{m-1}{\binom{2k}{k}\left(\frac{1-\mu^2\left(c\right)}{4}\right)^k}\right],
\\
\ &J\left(c\right)=\\
&\left[\frac{\Gamma\left(m+\frac{1}{2}\right)\sqrt \frac{c}{\pi}}{2\Gamma\left(m+1\right)\left(1+c\right)^{m+\frac{1}{2}}}{_2}F_1\left(1,m+\frac{1}{2};m+1;\frac{1}{1+c}\right)\right]
\end{split}
\end{equation}
şeklindedir. Burada $\mu\left(c\right)=\sqrt{\frac{c}{1+c}}$ olarak ifade edilir. $\Gamma\left(.\right)$ Gamma işlevi olarak adlandırılır ve $\Gamma\left(m\right)=\int_{0}^{\infty}{y^{m-1}e^{-y}dy}$ şeklinde tanımlanır. ${_2}F_1\left(a,b;c;z\right)$ ise Gauss Hipergeometrik işlevidir ve ${_2}F_1\left(a,b;c;z\right)=\sum_{k=0}^{\infty}\frac{\left(a\right)_k\left(b\right)_kz^k}{\left(c\right)_kk!}/;\ \left|z\right|<1\ \ \bigvee$ $\left|z\right|=1\ \bigwedge\ Re(c-a-b)>0$ olarak ifade edilir.

Yakın kullanıcıda gerçekleştirilen SIC işlemi sırasında öncelikle uzak kullanıcı sembolleri sezilecek ve sezilen bu uzak kullanıcı sembolleri alınan toplam işaretten çıkarılarak yakın kullanıcının kendi sembollerine karar verilecektir. Dolayısıyla yakın kullanıcıda iki farklı olayın meydana gelme ihtimali vardır. Bunlar; SIC sırasında uzak kullanıcı sembollerinin doğru sezilmesi ve hatalı sezilmesi durumlarıdır. Yakın kullanıcı BHO ifadesi de bu iki farklı durum göz önüne alındığında
 \begin{equation}
  \ P_1\left(e\right)=\ P_1\left(e\left|{correct}_2\right.\right)+P_1\left(e\left|{error}_2\right.\right)
 \end{equation}
olarak verilir. Burada $P_1\left(e\left|{correct}_2\right.\right)$ ve $P_1\left(e\left|{error}_2\right.\right)$ sırasıyla yakın kullanıcıda uzak kullanıcı sembollerinin doğru sezildiği ve hatalı sezildiği durumlardaki yakın kullanıcı için BHO ifadeleri olarak tanımlanır.
İlk olarak uzak kullanıcı sembollerinin doğru sezildiği durum incelenirse; uzak kullanıcı sembolleri için SIC işleminin ilk aşamasındaki karar kuralı, aynı uzak kullanıcıdaki karar kuralı gibi, $y_{1,R}<0$ veya $y_{1,R}\geq0$ olarak tanımlanır. Bu karar kuralı sonucunda karar verilen uzak kullanıcı sembollerinin doğru sezilme olasılığı yakın kullanıcıda SIC sırasında hata yapılmaması durumu için öncül olasılıklar olarak ifade edilir. Bu öncül olasılıkların farklı $x_1$ ve $x_2$ çiftleri için farklı olacağı Tablo I ve Denklem (2) den açıkça gözükmektedir. Sezilen uzak kullanıcı verileri toplam alınan işaretten çıkarıldıktan sonra yakın kullanıcı için ML karar kuralı $\acute{y_{1,R}}=0$ ve$\acute{\ y_{1,I}}=0$ eksenlerinin farklı bölgelerinde olması olarak tanımlanır. Burada $\acute{y_{1,R}}$ ve $\acute{\ y_{1,I}}$ sırasıyla, alınan $y_1$ işaretinden sezilen uzak kullanıcı sembollerinin çıkarılmasından sonra oluşan işaretin gerçel ve sanal kısımlarını göstermektedir. Yakın kullanıcıda uzak kullanıcı sembollerinin doğru sezilmesi şartı altında, öncül olasılıkları da içeren yakın kullanıcı sembollerinin BHO ifadesi
\begin{equation}
\begin{split}
&\ P_1\left(e\left|{correct}_2\right.\right)=\frac{1}{4}\left[P_r\left(n_R\le\sqrt{\varepsilon_2}h_1-\sqrt{\frac{\varepsilon_1}{2}}h_1\right)\right.\\
&\times\{P_r\left(n_R\le-\sqrt{\frac{\varepsilon_1}{2}}h_1\middle|\ n_R\le\sqrt{\varepsilon_2}h_1-\sqrt{\frac{\varepsilon_1}{2}}h_1\right)\\
&+P_r\left(n_I\geq\sqrt{\frac{\varepsilon_1}{2}}h_1\middle|\ n_R\le\sqrt{\varepsilon_2}h_1-\sqrt{\frac{\varepsilon_1}{2}}h_1\right)\}\\
&+P_r\left(n_R\le\sqrt{\varepsilon_2}h_1+\sqrt{\frac{\varepsilon_1}{2}}h_1\right)\\
&\times\{P_r\left(n_R\geq\sqrt{\frac{\varepsilon_1}{2}}h_1\middle|\ n_R\le\sqrt{\varepsilon_2}h_1+\sqrt{\frac{\varepsilon_1}{2}}h_1\right)\\
&\left.+P_r\left(n_R\geq\sqrt{\frac{\varepsilon_1}{2}}h_1\middle|\ n_R\le\sqrt{\varepsilon_2}h_1+\sqrt{\frac{\varepsilon_1}{2}}h_1\right)\}\right]
\end{split}
\end{equation}
olarak bulunur. Burada $n_R$ ve $n_I$ ifadeleri AWGN’nin gerçel ve sanal bileşenlerini göstermektedir. Denklem (7)'de öncül olasılıklar ($0.5P_r\left(n_R\le\sqrt{\varepsilon_2}h_1\pm\sqrt{\sfrac{\varepsilon_1}{2}}h_1\right)$) Denklem (2)’ye benzer şekilde yakın kullanıcıda uzak kullanıcı sembollerinin doğru çözülme olasılıkları olarak ifade edilir. Koşullu olasılık ifadeleri ise uzak kullanıcı sembollerine doğru karar verilmesi koşulu altındaki hata olasılıklarını göstermektedir. Verilen koşullu olasılık ifadeleri yerine konup ve birkaç cebirsel işlemden sonra, uzak kullanıcı sembollerinin doğru çözülmesi şartı altında yakın kullanıcı için BHO ifadesi
\begin{equation}
\begin{split}
&P_1\left(e\left|{correct}_2\right.\right)=\\
&\frac{1}{4}\left[Q\left(\sqrt{\gamma_C}\right)\times\left\{4-Q\left(\sqrt{\gamma_D}\right)-Q\left(\sqrt{\gamma_E}\right)\right\}-Q\left(\sqrt{\gamma_D}\right)\right]
\end{split}
\end{equation}
şeklinde elde edilir. Burada yine gösterim kolaylığı açısından $\gamma_C=\frac{\varepsilon_1\times\left|h_1\right|^2}{N_0}$, $\gamma_D=\frac{\left(\sqrt{2\varepsilon_2}+\sqrt{\varepsilon_1}\right)^2\times\left|h_1\right|^2}{N_0}$ ve $\gamma_E=\frac{\left(\sqrt{2\varepsilon_2}-\sqrt{\varepsilon_1}\right)^2\times\left|h_1\right|^2}{N_0}$ tanımlamaları yapılmıştır.

Yakın kullanıcıda gerçekleştirilen SIC işlemi sırasında uzak kullanıcı sembollerinin hatalı sezildiği durumda; uzak kullanıcı ve yakın kullanıcı sembolleri için ML karar kurallarının uzak kullanıcı sembollerinin doğru çözüldüğü durumdaki gibi olacağı açıktır. Fakat, buradaki öncül olasılıklar ve koşul ifadeleri uzak kullanıcı sembollerinin hatalı sezilme olayına bağlı olarak değişecektir. Ayrıca elde edilen $\acute{y_1}$ işaretinin de işaret uzayının farklı olacağı açıktır. Tüm bu durumlar göz önünde bulundurulduğunda Denklem (7)'ye benzer olarak, SIC sırasında uzak kullanıcı sembollerinin hatalı çözülmesi koşulu altında yakın kullanıcı sembolleri için BHO ifadesi
\begin{equation}
\begin{split}
&P_1\left(e\left|{error}_{2}\right.\right)=\frac{1}{4}\left[P_r\left(n_R\geq\sqrt{\varepsilon_2}h_1-\sqrt{\frac{\varepsilon_1}{2}}h_1\right)\right.\\
&\times\{P_r\left(n_R\le2\sqrt{\varepsilon_2}h_1-\sqrt{\frac{\varepsilon_1}{2}}h_1\middle|\ n_R\geq\sqrt{\varepsilon_2}h_1-\sqrt{\frac{\varepsilon_1}{2}}h_1\right)\\
&+P_r\left(n_I\geq\sqrt{\frac{\varepsilon_1}{2}}h_1\middle|\ n_R\geq\sqrt{\varepsilon_2}h_1-\sqrt{\frac{\varepsilon_1}{2}}h_1\right)\}\\
&+P_r\left(n_R\geq\sqrt{\varepsilon_2}h_1+\sqrt{\frac{\varepsilon_1}{2}}h_1\right)\\
&\times\{P_r\left(n_R\geq2\sqrt{\varepsilon_2}h_1+\sqrt{\frac{\varepsilon_1}{2}}h_1\middle|\ n_R\geq\sqrt{\varepsilon_2}h_1+\sqrt{\frac{\varepsilon_1}{2}}h_1\right)\\
&\left.+P_r\left(n_I\geq\sqrt{\frac{\varepsilon_1}{2}}h_1\middle|\ n_R\geq\sqrt{\varepsilon_2}h_1+\sqrt{\frac{\varepsilon_1}{2}}h_1\right)\}\right]
\end{split}
\end{equation}
olarak bulunur. Bir önceki duruma benzer şekilde Denklem (9)’daki öncül olasılıklar SIC sırasında uzak kullanıcı sembollerinin hatalı çözüme olasılıklarını göstermekte ve koşul ifadeleri SIC sırasında uzak kullanıcı sembollerine hatalı karar verilmesi olayını ifade etmektedir.  Bir kaç cebirsel işlemden sonra, SIC sırasında uzak kullanıcı sembollerinin hatalı sezilmesi şartı altında, yakın kullanıcı sembolleri için BHO ifadesi 
\begin{equation}
\begin{split}
&P_1\left(e\left|{error}_2\right.\right)=\frac{1}{4}\left[Q\left(\sqrt{\gamma_C}\right)\times \right.\\
&\left.\{Q\left(\sqrt{\gamma_D}\right)+Q\left(\sqrt{\gamma_E}\right)\}+Q\left(\sqrt{\gamma_E}\right)+Q\left(\sqrt{\gamma_F}\right)-Q\left(\sqrt{\gamma_G}\right)\right]
\end{split}
\end{equation}
şeklinde ifade edilir. Aynı gösterimi kullanmak adına $\gamma_F=\frac{\left(2\sqrt{{2\varepsilon}_2}+\sqrt{\varepsilon_1}\right)^2\times\left|h_1\right|^2}{N_0}$ ve $\gamma_G=\frac{\left(2\sqrt{{2\varepsilon}_2}-\sqrt{\varepsilon_1}\right)^2\times\left|h_1\right|^2}{N_0}$ olarak tanımlanmıştır. Denklem (8) ve (10)'da verilen ifadeler Denklem (6)'da yerine konulduktan sonra; Denklem (3)'de olduğu gibi [12-14]‘de verilen ifadelerden yararlanılarak yakın kullanıcı için ortalama BHO ifadesi Denklem (5)'te tanımlanan işlevler cinsinden
 \begin{equation}
\begin{split}
  &\overline{P_1}\left(e\right)= \\
	&\begin{cases}
         I\left(c_C\right)+\frac{1}{4}\left[-I\left(c_D\right)+I\left(c_E\right)+I\left(c_F\right)-I\left(c_G\right)\right] \text{$m$ tamsayı,}
        \\
        J\left(c_C\right)+\frac{1}{4}\left[-J\left(c_D\right)+J\left(c_E\right)+J\left(c_F\right)-J\left(c_G\right)\right] \text{diğer}.
        \end{cases}
				\end{split}
 \end{equation}
şeklinde elde edilir. Burada $c_i=\frac{\overline{\gamma_i}}{2m},\ i=C,D...G.$, $\overline{\gamma_C}=E\left[\gamma_C\right]$, $\overline{\gamma_D}=E\left[\gamma_D\right]$, $\overline{\gamma_E}= E\left[\gamma_E\right]$, $\overline{\gamma_F}= E\left[\gamma_F\right]$ ve $\overline{\gamma_G}=E\left[\gamma_G\right]$ olarak tanımlanır.
\section{ Nümer{\footnotesize İ}k Sonuçlar}
Bu bölümde bir önceki bölümde türetilen analitik BHO ifadelerinin bilgisayar benzetimleri ile doğrulamaları sunulmuştur. Monte Carlo benzetimleri ile elde edilen analitik sonuçların birebir örtüştüğü gözlenmiştir.

Şekil 1’de $m$ parametresinin sistemin BHO performansına etkisi incelenmiştir. Şekil 1a’da yakın kullanıcı için $m$ parametresinin etkisi sunulmuştur. $m_1=1,2,3,4$ olarak alınmıştır. Şekil 1b’ de uzak kullanıcı için $m$ parametresinin etkisi gösterilmiştir. $m_2=0.5,1,1.5,2$ değerleri için benzetim ve analitik sonuçlar sunulmuştur. Her iki senaryoda da ortalama kanal güçleri $\Omega_1=\Omega_2=0dB$, ve güç paylaşım katsayısı $\alpha=0.3$ olarak seçilmiştir. NOMA tekniğinde $m$ parametresinin her iki kullanıcı için çeşitlilik (diversity) derecesini arttırdığı gözlenmiştir. 
\begin{figure}[!t]
	\centering
	\shorthandoff{=}  %\usepackage[turkish]{babel} kullanıldığı için komuta gerek var
	\includegraphics[height=11cm, width=7cm]{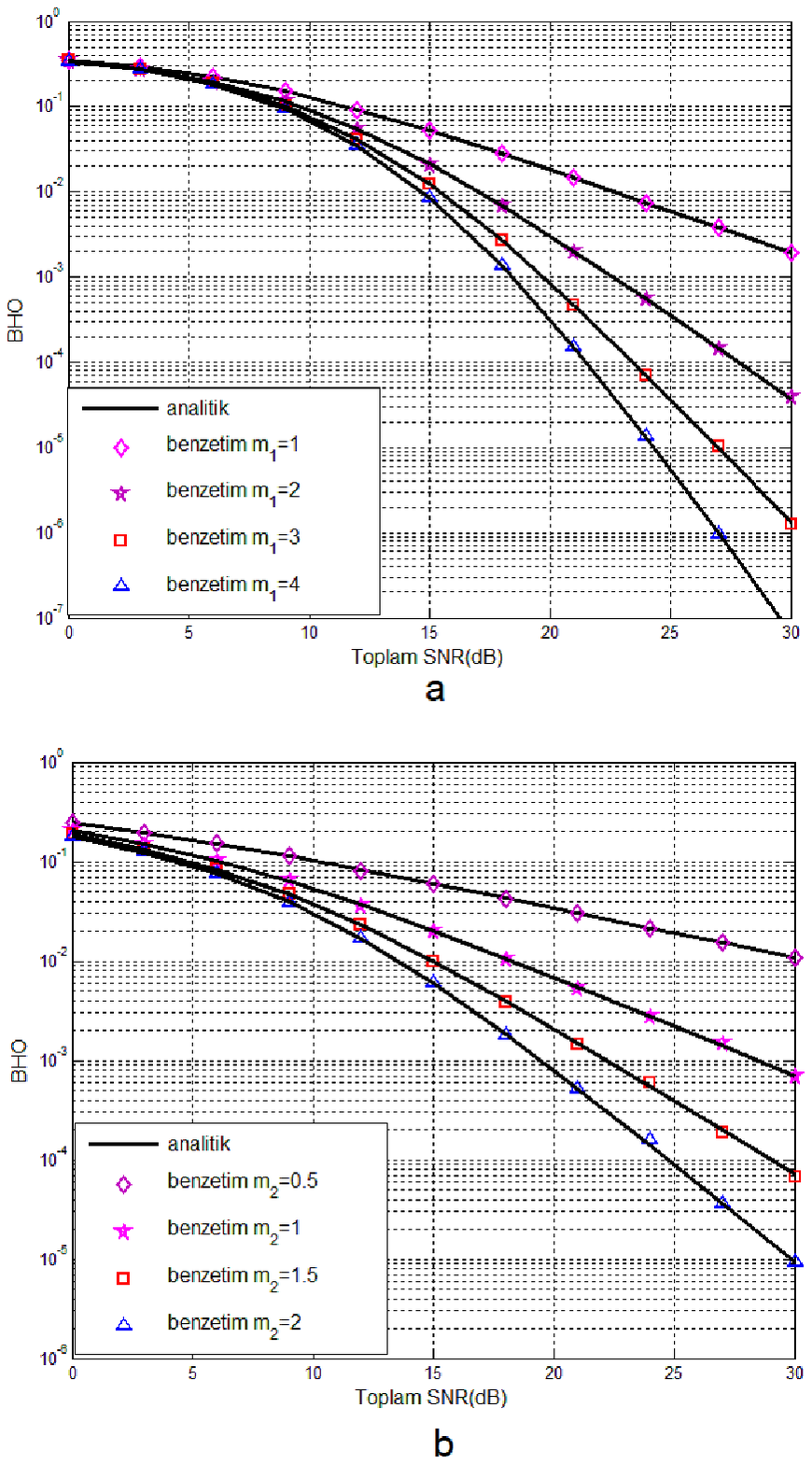}
	\shorthandon{=} %\usepackage[turkish]{babel} kullanıldığı için komuta gerek var
	\caption{\footnotesize DL-NOMA Kullanıcılarının BHO Performanları a. Yakın Kullanıcı BHO Performansı $m_1=1,2,3,4$ b.Uzak Kullanıcı BHO Performansı $m_2=0.5,1,1.5,2$}
	\label{sekil1}
\end{figure}

Şekil 2’de kullanıcıların BHO performansları $\alpha$ değerinin değişimine göre verilmiştir. Sonuçlar $m_1=1,2,3,4$ ve verici $SNR=20dB$ için sunulmuştur. $\alpha$ değerinin artması uzak kullanıcı sembollerine aktarılan gücü azalttığından uzak kullanıcı BHO performansının düşeceği açıktır. Fakat, $\alpha$ değerinin artması yakın kullanıcı BHO performansının sürekli olarak artacağı anlamına gelmemektedir. Bu durum $\alpha$ değerinin çok artması durumunda yakın kullanıcıda gerçekleştirilen SIC sırasında uzak kullanıcı sembollerinin hatalı sezilmesini arttırmaktadır ve bu da yakın kullanıcı için BHO performansını negatif yönde etkilemektedir.
\begin{figure}[!t]
	\centering
	\shorthandoff{=}  %\usepackage[turkish]{babel} kullanıldığı için komuta gerek var
\includegraphics[height=5cm, width=7.5cm]{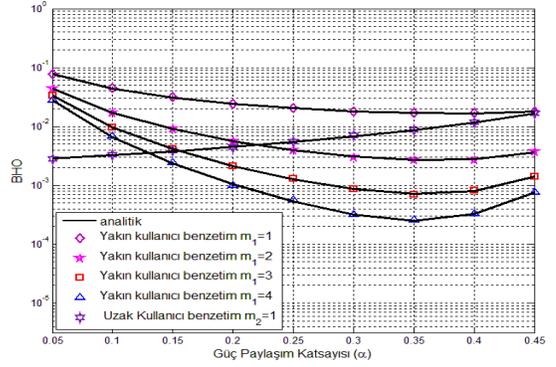}
	\shorthandon{=} %\usepackage[turkish]{babel} kullanıldığı için komuta gerek var
	\caption{\footnotesize Güç Paylaşım Katsayısının BHO Performansına Etkisi}
	\label{sekil1}
\end{figure}
\section{Sonuçlar ve Tartışma}
Bu çalışmada literatürde ilk defa aşağı yönlü NOMA sisteminin Nakagami-$m$ sönümlemeli kanallardaki BHO ifadeleri SIC hataları varlığında kapalı formda türetilmiştir. Elde edilen ifadeler bilgisayar benzetimleri ile doğrulanmıştır. NOMA tekniği kullanılan Tek Girişli Tek Çıkışlı (Single Input Single Output -SISO) sistemler için $m$ parametresi çeşitlilik derecesini ifade etmektedir.
Benzetim sonuçlarından da görüleceği gibi $\alpha$ güç paylaşım katsayısı değeri sistem BHO performansları üzerinde önemli bir etkiye sahiptir. Özellikle yakın kullanıcıda gerçeklenen SIC işleminde yapılan hataya etkisi nedeniyle $\alpha$ değerinin belirlenmesi son derece önemlidir. 
\section{B{\footnotesize İ}lg{\footnotesize İ}lend{\footnotesize İ}rme}
Bu  çalışma TÜBİTAK tarafından 2211-E Doktora Bursu kapsamında desteklenmektedir.

\end{document}